\documentclass{appolb}%
\usepackage{amsmath,amsfonts,amssymb,amscd,amsxtra,amsthm}
\usepackage{bm}
\usepackage{multirow}
\usepackage{graphicx}
\usepackage{color}%
\usepackage{amsfonts}

\begin{document}
\renewcommand*{\thefootnote}{\fnsymbol{footnote}}

\title{On a possibility of baryonic exotica\thanks{Presented 
at the 
{\em 2nd Jagiellonian Symposium on Fundamental and Applied Subatomic Physics}, June 3-11, 2017, Krak{\'o}w, Poland.}}
\author{Micha{\l} Prasza{\l}owicz
\address{M. Smoluchowski Institute of Physics, Jagiellonian University, \\
ul. S. {\L}ojasiewicza 11, 30-348 Krak{\'o}w, Poland.}}

\maketitle

\begin{abstract}
Models based on chiral symmetry predict pentaquarks that have relatively low masses. 
We briefly review both theoretical 
and experimental status of exotica in the light sector. Next,  shall show how to extend chiral
models to baryons with one heavy quark and show that one expects exotica
also in this case. Finally, we  interpret recently discovered
by the LHCb Collaboration five $\mathit{\Omega}^*_c$ resonances in terms of regular and exotic
excitations of the ground state $\mathit{\Omega}_c$.
\end{abstract}


\bigskip
\bigskip


\section{Introduction: Chiral Quark-Soliton Model}
\label{sec:intro}

\begin{figure}[h]
\centering
\includegraphics[width=7.9cm]{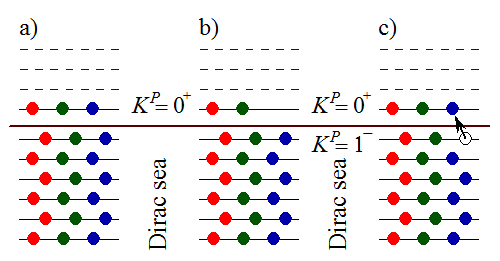} \vspace{-0.2cm}\caption{Schematic
pattern of light ($u$ and $d$) quark levels in a self-consistent soliton
configuration. In the left panel all sea levels are filled and $N_{c}$ (=3 in
the figure) valence quarks occupy the $K^{P}=0^{+}$ lowest
positive energy level.  In the middle panel, one valence quark has been
stripped off, and the soliton has to be 
supplemented by a heavy quark not shown in the figure. In the right panel, a
possible excitation of a sea level quark, conjectured to be $K^{P}=1^{-}$, to
the valence level is shown, and again the soliton has to couple to a heavy
quark. Levels for strange quarks that exhibit different filling pattern are
not shown.}%
\label{fig:levels}%
\end{figure}

The Chiral Quark-Soliton Model ($\chi$QSM) is based on an old argument by Witten
\cite{Witten:1979kh}, which says that in the  $N_c \rightarrow \infty$   limit  ($N_c$ stands for number of colors),
$N_{c}$ relativistic valence quarks generate chiral mean fields
represented by a distortion of a Dirac sea that in turn 
interact with the valence quarks themselves (for  a review see
Ref.\cite{Christov:1995vm}).  In this way, a self-consistent configuration called a
\emph{soliton} is formed. In Fig.~\ref{fig:levels}~(a) we plot  schematic pattern of light quark energy levels corresponding
to this scenario. It is assumed that the mean fields exhibit so called {\em hedgehog} symmetry, which means that 
neither quark spin ($\bm{S}_q$) nor quark isospin ($\bm{T}_q$)  are "good" quantum numbers. Instead a {\em grand spin}  
$\bm{K}=\bm{S}_q+\bm{T}_q$ is a "good" quantum number. The lowest valence level has $K^P=0^+$. 

In order to project
out spin and isospin one has to rotate the
the soliton, both in flavor and configuration spaces. These rotations are then quantized
semiclassically and the collective Hamiltonian  is computed. 
The model
predicts rotational 
baryon spectra that satisfy the following selection rules:

\begin{itemize}
\item allowed SU(3) representations must contain states with hypercharge
$Y^{\prime}=N_{c}/3$,
\vspace{-0.25cm}
\item the isospin $\bm{T}^{\prime}$ of the states with $Y^{\prime}%
=N_{c}/3$ couples with the soliton spin $\bm{J}$ to a singlet:
$\bm{T}^{\prime}+\bm{J}=0$.
\end{itemize}

\begin{figure}[h!]
\centering
\includegraphics[width=10cm]{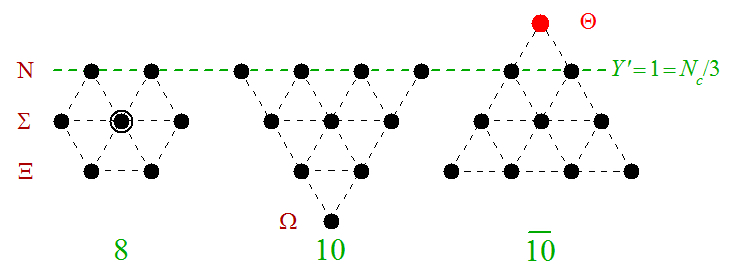} \vspace{-0.1cm}
\caption{Lowest lying SU(3) flavor representations allowed by the constraint $Y'=1$. The first 
{\em exotic} representation, $\overline{\mathbf{10}}$, contains the explicitly
exotic pentaquark states $\mathit{\Theta}^+$,
$\mathit{\Xi}^{+}$  and $\mathit{\Xi}^{--}$ and non-exotic nucleon- and sigma-like states.}
\label{fig:irreps1}%
\end{figure}

In the case of light positive parity baryons the lowest allowed representations are
$\mathbf{8}$ of spin 1/2, $\mathbf{10}$ of spin 3/2, and also exotic
$\overline{\mathbf{10}}$ of spin 1/2 with the lightest state corresponding to
the putative $\mathit{\Theta}^{+}(1540)$. They are shown in Fig.~\ref{fig:irreps1}. 
Chiral models in general predict that pentaquarks are light 
\cite{Praszalowicz:2003ik,Diakonov:1997mm}
and -- in some specific models -- narrow \cite{Diakonov:1997mm}.

After 
the first enthusiastic announcements of the discovery of pentaquarks in 2003 
by LEPS
\cite{Nakano:2003qx} and DIANA \cite{Barmin:2003vv} collaborations,
the experimental evidence for the light
exotica has been questioned (see {\em e.g.} \cite{Hicks:2012zz}). Nevertheless,
both DIANA \cite{Barmin:2013lva} and LEPS \cite{Nakano:2017fui} upheld their 
original claims after performing higher statistics analyses. The report on
exotic $\mathit{\Xi}$ states (see Fig.~\ref{fig:irreps1}) by NA49 \cite{Alt:2003vb}
from 2004, to the best of my knowledge, has not been questioned so far, 
however the confirmation is still strongly needed.

Another piece of information on $\overline{\mathbf{10}}$ comes from the
$\eta$ photo-production off the nucleon.  Different experiments confirm the narrow
structure at the c.m.s. energy $W \sim  1.68$~GeV
observed in the case of the neutron, whereas no structure is observed on the
proton (see Fig.~27 in the latest report by CBELSA/TAPS Collaboration \cite{Witthauer:2017pcy} 
and  references therein). The natural interpretation of this "neutron puzzle" was proposed already
in 2003 in Ref.~\cite{Polyakov:2003dx}.
There one assumes that the narrow excitation at $W \sim  1.68$~GeV corresponds to the non-exotic
penta-nucleon resonance belonging to $\overline{\mathbf{10}}$.
Indeed, the SU(3) symmetry forbids photo-excitation of the proton member of $\overline{\mathbf{10}}$, while
the analogous transition on the neutron is possible. This is due to the fact that photon is an SU(3)  $U$-spin singlet,
and the $U$-spin symmetry is exact in the SU(3) symmetric limit. An alternative interpretation is based
on a partial wave analysis in terms of  the Bonn-Gatchina approach \cite{Anisovich:2017xqg}. 
There is an ongoing dispute on the interpretation of the "neutron puzzle" (for the latest arguments
see Ref.~\cite{Kuznetsov:2017qmo}).

\section{Heavy Baryons in the Chiral Quark-Soliton Model}
\label{sec:HB}

\begin{figure}[h!]
\centering
\includegraphics[width=9cm]{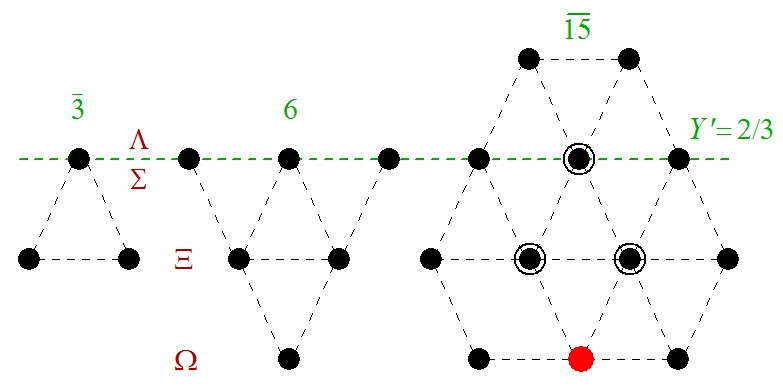} \vspace{-0.1cm}
\caption{Lowest lying SU(3) flavor representations allowed by the constraint $Y'=2/3$. The first 
{\em exotic} representation, $\overline{\mathbf{15}}$, contains the putative pentaquark states $\mathit{\Omega}_c$
with $\mathit{\Omega}_c^0$ marked in red.}
\label{fig:irreps2}%
\end{figure}

In a recent paper \cite{Yang:2016qdz} following \cite{Diakonov:2010tf} we
have extended the $\chi$QSM to baryons involving one heavy quark. In this case
the valence level is occupied by $N_{c}-1$ light quarks (see
Fig~\ref{fig:levels}~(b)) that couple with a heavy quark $Q$ to form a color
singlet. The lowest allowed SU(3) representations are shown in Fig.~\ref{fig:irreps2}.
They correspond to the soliton in representation in $\overline{\mathbf{3}}$ of spin 0  and 
to ${\mathbf{6}}$ of spin 1. Therefore, the baryons constructed  from such a soliton and a heavy
quark form an SU(3) antitriplet of spin 1/2 and two sextets  of spin 1/2 and 3/2 that are subject to a hyper-fine
splitting. The next allowed representation of the rotational excitations
corresponds to the exotic $\overline{\mathbf{15}}$ of spin 0 or spin
1 \cite{Kim:2017jpx}. The spin 1 soliton has  lower mass
and when it couples with a heavy quark, it forms spin 1/2 or 3/2 exotic 
multiplets that should be hyper-fine split similarly to the ground state sextets by
$\sim 70$~MeV.

The rotational states described above  correspond to positive
parity and are clearly seen in the data \cite{Kim:2017jpx}. Negative parity states are generated
by soliton configurations with one light  quark excited to
the valence level from the Dirac sea (Fig.~\ref{fig:levels}~(c)).  The
selection rules for excited quark solitons can be  summarized as follows \cite{Petrov:2016vvl}:
\begin{itemize}
\item allowed SU(3) representations must contain states with hypercharge
$Y^{\prime}=(N_{c}-1)/3$,
\vspace{-0.25cm}
\item the isospin $\bm{T}^{\prime}$ of the states with $Y^{\prime}%
=(N_{c}-1)/3$ couples with the soliton spin $\bm{J}$ as follows:
$\bm{T}^{\prime}+\bm{J}=\bm{K}$, where $\bm{K}$ is the grand spin
of the excited level.
\end{itemize}

The first allowed SU(3) representation for one quark excited soliton is again
$\overline{\mathbf{3}}$, Fig.~\ref{fig:irreps2}, with $T^{\prime}=0$, which
 for $K=1$ is quantized as spin 1. The coupling
of a heavy quark results in two hyperfine split antitriplets that are indeed seen
in the data \cite{Kim:2017jpx}. The hyperfine splitting parameter is in this case 
 $\kappa^{\prime}/m_{c}\sim 30$~MeV.
Next possibility is flavor $\mathbf{6}$ with
$T^{\prime}=1$, which may couple with $K=1$ to $J=0,1$ and $2$
resulting in 5 hyperfine split heavy sextets: two $1/2^-$, two $3/2^-$
and one $5/2^-$ (see Tab.~\ref{tab:s1}).

\section{Possible interpretation of the LHCb $\mathit{\Omega}_c^0$ resonances}
\label{sec:LHCb}

In a very recent paper the LHCb Collaboration announced five
$\mathit{\Omega}^{0}_{c}$ states with masses in the range of $3 - 3.2$~GeV
\cite{Aaij:2017nav}. The simplest
possibility would be to associate them with the five sextets described 
at the end of Sect.~\ref{sec:HB}.
We have shown, however, in \cite{Kim:2017jpx} that this scenario fails, as can be seen from 
Table~\ref{tab:s1}.

\renewcommand{\arraystretch}{1.3}
\begin{table}[th]
\begin{center}%
\begin{tabular}
[c]{|c|c|c|c|c|}\hline
$J$ & $S^{P}$ & $M$~[MeV] & $\kappa^{\prime}/m_{c}$~[MeV] & $\Delta_{J}$~[MeV]\\ \hline
0 & ${1}/{2}^{-}$ & 3000 & -- & --\\\hline
\multirow{2}{*}{1} & ${1}/{2}^{-}$ & 3050 & \multirow{2}{*}{16} &
\multirow{2}{*}{61}\\
~ & ${3}/{2}^{-}$ & 3066 &  & \\\cline{1-5}%
\multirow{2}{*}{2} & ${3}/{2}^{-}$ & 3090 & \multirow{2}{*}{17} &
\multirow{2}{*}{47}\\
& ${5}/{2}^{-}$ & 3119 &  & \\\hline
\end{tabular}
\end{center}
\par
\vspace{-0.3cm} \vspace{0.3cm}\caption{$\chi$QSM scenario where all LHCb $\mathit{\Omega}_{c}^0$
states are assigned to the excited sextets. This assignment requires hyperfine
splitting which is almost two times smaller than in the $\overline{\mathbf{3}%
}$ case and relation $\Delta_2=2 \Delta_1$ derived in \cite{Kim:2017jpx} is badly broken.\
Here $\Delta_J$ is the mass difference between states of given $J$ and $J-1$ before hyper-fine splitting.}%
\label{tab:s1}%
\end{table}
\renewcommand{\arraystretch}{1}

\begin{figure}[h!]
\centering
\includegraphics[width=9cm]{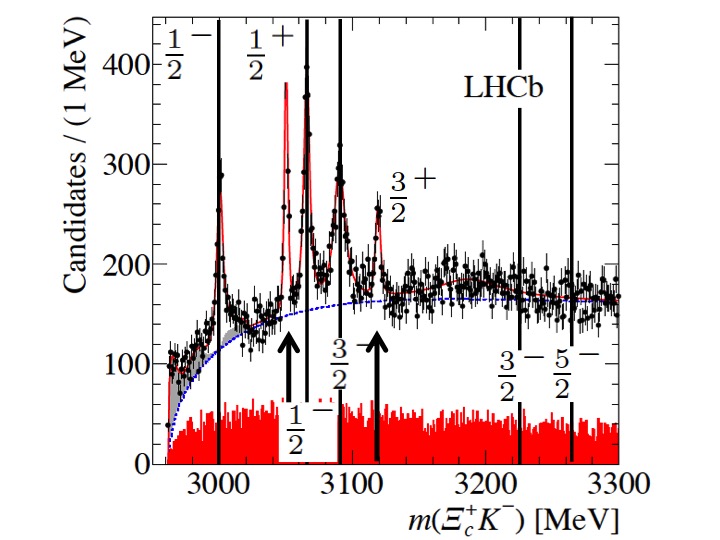} \vspace{-0.1cm}
\caption{LHCb spectrum \cite{Aaij:2017nav} with the assignment described in the text. Two narrow 
$1/2^+$ and $3/2^+$ states marked with arrows are interpreted as $\overline{\mathbf{15}}$ pentaquarks.}
\label{fig:spectrum}%
\end{figure}

In the second scenario proposed in \cite{Kim:2017jpx}, we have interpreted three LHCb states as quark
excitations of the ground state sextets, shown in Fig.~\ref{fig:spectrum} as vertical lines. Two remaining sextet 
excitations have higher mass and are above the threshold for the decays into charm mesons. They can be, therefore,
wide and the branching ratio to $\mathit{\Xi}^+_c +\mathit{K}^-$ final state may be small. This would explain why they are not
seen by LHCb. On the other hand, two remaining $\mathit{\Omega}_c^0$ peaks are in this scenario interpreted as rotational
excitations corresponding to the exotic $\overline{\mathbf{15}}$. As such, they are isospin triplets and should decay 
not only to $\mathit{\Xi}^+_c + \mathit{K}^-$ but also to
 $\mathit{\Xi}_c^0+\mathit{K}^-$ or $\mathit{\Xi}_c^+ + \bar{ \mathit{K}}^0$  
 and $\mathit{\Omega}_c+\mathit{\pi}$ final states. This scenario is,
 therefore, very easy to confirm or falsify. Moreover, they are very narrow with widths around 1~MeV,
 and the $\chi$QSM provides a mechanism that suppresses pentaquark decays both in the light sector and
 in the present approach to heavy baryons \cite{Diakonov:1997mm,Kim:2017jpx}.
 
 \vspace{0.1cm}
 
 Summarizing, let us stress that despite many "null findings" there is still an experimental support for
 light and narrow pentaquarks. Using the ideas of the $\chi$QSM, we have proposed an interpretation
 of recently discovered $\mathit{\Omega}_c^0$ states in terms of quark and and rotational excitations of the
 ground state charmed baryons, the latter corresponding to the pentaquarks.

\section*{Acknowledgemens}
This note is based on Refs.~\cite{Yang:2016qdz,Kim:2017jpx} where more complete list of references can be found.
I would like to thank H.C. Kim, M.V. Polyakov and G.S. Yang for a fruitful collaboration.

\end{document}